\documentclass{emulateapj}
\lefthead{GAUDI \& HAN} \righthead{MICROLENSING EVENT OGLE-2002-BLG-055}

\def\eq#1{equation (\ref{#1})}
\def\Eq#1{Eq.~\ref{#1}}
\def\te{t_{\rm E}}
\def\cax{{\cal A}_{x}}
\def\cay{{\cal A}_{y}}
\def\thetae{\theta_{\rm E}}
\def\dos{D_{\rm os}}

\def\dol{D_{\rm ol}}
\def\fs{F_s}
\def\fb{F_b}
\def\event{OGLE-2002-BLG-055}
\def\pirel{\pi_{\rm rel}}
\def\murel{\mu_{\rm rel}}
\def\rehat{\hat r_{\rm E}}

\begin{document}

\title{The Many Possible Interpretations of Microlensing Event
OGLE-2002-BLG-055} \author {B.\ Scott Gaudi\altaffilmark{1} and
Cheongho Han\altaffilmark{2,3}} \altaffiltext{1}{Harvard-Smithsonian
Center for Astrophysics, 60 Garden St., Cambridge, MA 02138}
\altaffiltext{2}{Department of Physics, Institute for Basic Science
Research, Chungbuk National University, Chongju 361-763, Korea}
\altaffiltext{3}{Department of Astronomy, Ohio State University,
140 West 18th Ave., Columbus, OH 43210}
\email{sgaudi@cfa.harvard.edu,cheongho@astroph.chungbuk.ac.kr}

\begin{abstract}

Microlensing event \event\ is characterized by a smooth, slightly
asymmetric single-lens curve with an isolated, secure data point that is
$\sim 0.6$ magnitudes brighter than neighboring points separated by a
few days.  It was previously suggested that the single deviant data
point and global asymmetry were best explained by a planetary
companion to the primary lens with mass ratio $q=10^{-3}-10^{-2}$, and
parallax effects induced by the motion of the Earth.  We revisit the
interpretation of \event, and show that the data can be explained by
wide variety of models.  We find that the deviant data point can be
fit by a large number of qualitatively different binary-lens models
whose mass ratios range, at the $\sim 3\sigma$ level, from $q\simeq
10^{-4}$ to $\simeq 10^{-1}$.  This range is consistent with a planet,
brown dwarf, or M-dwarf companion for reasonable primary masses of
$M\ga 0.8M_\odot$.  A subset of these binary-lens fits consist of a
family of continuously degenerate models whose mass ratios differ by
an order-of-magnitude, but whose light curves differ by $\la 2\%$ for
the majority of the perturbation.  The deviant data point can also be
explained by a binary companion to the {\it source} with
secondary/primary flux ratio of $\sim 1\%$.  This model has the added
appeal that the global asymmetry is naturally explained by the
acceleration of the primary induced by the secondary.  
The binary-source model yields a measurement of the
Einstein ring radius projected onto the source plane of ${\hat r_{\rm
E}}=1.87\pm 0.40~{\rm AU}$.  \event\ is an extreme example that
illustrates the difficulties and degeneracies inherent in the
interpretation of weakly perturbed and/or poorly sampled microlensing
light curves.

\end{abstract}
\keywords{gravitational lensing -- planetary systems}

\section{Introduction\label{sec:intro}}

Planetary companions to Galactic disk and bulge microlens stars can be
discovered via the short duration perturbation they create to the
smooth light curve induced by the parent star \citep{mp91}.  The
majority of these perturbations are expected to be relatively simple
and grossly characterized by three observables: the duration, peak
time, and magnitude of the perturbation.  In the ideal scenario, these
three observables are simply related to the three parameters
describing the planetary system: the planet/star mass ratio, the
instantaneous projected separation in units of the angular Einstein
ring radius, and the angle of the source trajectory relative to the
planet/star axis \citep{gl92}.  Unfortunately, reality is a bit more
complicated, and a number of degeneracies have been identified which
can hamper the ability to infer these parameters in practice.
\citet{gandg97} demonstrated that there exists an ambiguity in the
physical mechanism that sets the width of low-amplitude perturbations
which can result in an order-of-magnitude uncertainty in the inferred
mass ratio.  \citet{gaudi98} pointed out that a subset of binary
sources with extreme flux ratios can reproduce the duration and
magnitude of a subset of planetary microlensing perturbations,
although \citet{han02} demonstrated that this degeneracy could be
resolved with astrometric observations during the perturbation.
\citet{gs98} discuss a two-fold discrete degeneracy in the projected
separation prevalent in high-magnification planetary events.  Along
with these anticipated degeneracies, a few have been uncovered in the
process of detailed modeling of observed events.  \citet{bennett99}
invoked a planetary companion to explain a short-duration deviation
seen on a close binary-lens light curve.  However, it was later shown
by \citet{albrow00} that this perturbation could also be fit by one of
the secondary caustics of the close binary lens, when rotation of the binary
is considered.
\citet{gaudi02} found a weakly asymmetric event that could be equally
well-explained by a planetary companion, or parallax deviations
arising from the motion of the Earth.  Many of these degeneracies are
`accidental,' in the sense that they arise from chance similarities
between deviations caused by different physical situations,
rather than by intrinsic degeneracies in the lens equation itself. They are
therefore generally only approximate degeneracies, and can be resolved
with accurate, well-sampled light curves.  Given the short duration
and unpredictability of planetary deviations, dense, continuous and
accurate light curve coverage is necessary to both detect and
accurately characterize planetary microlensing perturbations.

A lensing star with a planetary companion is just an extreme limit of
a binary-lens.  As discussed by numerous authors, binary lenses are
themselves subject to numerous degeneracies
\citep{md95,albrow99,jm01}, some of which are rooted in symmetries in
the lens equation itself \citep{dominik99}, and are therefore nearly
perfect \citep{afonso00,albrow02}.  Binary lenses in which the source
does not cross any caustics can also be confused with binary sources.
This is especially problematic when only single-band photometry is
available.  This may partially account for the fact that, although
they were predicted to be plentiful \citep{gh92}, only one candidate
binary-source lensing event has been identified\footnote{See
\citet{dominik98} and \citet{hj98} for additional discussions of the
apparent lack of binary-source events.}, event OGLE-2003-BLG-095
\citep{collinge04}.  Indeed, \citet{collinge04} found that the
binary-source model for OGLE-2003-BLG-095 is only preferred over a
binary-lens model at the $\sim 3\sigma$ level. 

Source and lens binarity are not the only regimes where
degeneracies are plentiful; global deviations from the fiducial
point-source, point-lens, uniform motion (i.e.\ \citealt{pac86}) light
curve have also been found to be subject to degeneracies.  Such 
deviations come in a variety of forms.  The motion of the Earth 
produces departures from uniform relative motion which can induce
observable deviations from the
standard lightcurve form.  These deviations can be quite dramatic for events with
timescales $\te$ of order or larger than a year \citep{smith02}.
However, in the more usual case where $\te\ll {\rm yr}$, the effect of
the motion of the Earth can be approximated by a constant
acceleration, which results in deviations that can be
symmetric or asymmetric with respect to the peak of the event
\citep{gmh94,smp03}, although asymmetric deviations are generally
easier to recognize.  Unfortunately, as demonstrated by \citet{smp03},
any such weak parallax deviation can also be explained by acceleration of the
source, thus making the parallax interpretation non-unique for such
short timescale events. \citet{smp03} furthermore demonstrated that
constant-acceleration events are subject to a two-fold discrete
degeneracy between the magnitude and direction of the acceleration and
the event timescale.  \citet{gould03} showed that, in fact, light
curve degeneracies extend to even higher order, and are present when
one takes into account not only the acceleration, but also the jerk.  This
jerk-parallax degeneracy has been seen in one event toward the bulge
\citep{park04}, and has been invoked to resolve the discrepancy
between the photometric and microlensing mass determinations of the
microlens in the Large Magellanic Cloud event MACHO-LMC-5
\citep{gould03}.

Galactic bulge microlensing event \event\ exhibits both a global
asymmetry and single deviant data point.  \citet{jp02} first
considered the interpretation of this event; they fit binary-lens
models and included non-uniform motion caused by either parallax or
arbitrary uniform acceleration.  They argued that, although both
uniform acceleration and parallax explain the global deviations
equally well, the event was most naturally explained by
parallax-induced deviations, and that the single deviant point was
best fit by a binary-lens with mass ratio $10^{-3}-10^{-2}$.  This
mass ratio implies a companion in the Jupiter mass range for likely
primary masses, and they therefore concluded that \event\ was a
possible planet candidate.  Here we revisit the interpretation of this
event.  We demonstrate that, in fact, there are many possible
interpretations of \event.  We find that the short-timescale perturbation
could arise from a stellar companion to the source, or a stellar,
brown-dwarf, or planetary companion to the lens.  The global asymmetry
could arise from parallax deviations, or from acceleration of the
source induced by its companion.  All of these interpretations are
indistinguishable at the $\sim 3\sigma$ level.  Therefore, the correct
interpretation of this event is unclear, and it serves as a
particularly extreme reminder of the degeneracies involved with poorly
sampled and weakly-perturbed microlensing light curves.

\begin{figure}
\epsscale{1.1} 
\plotone{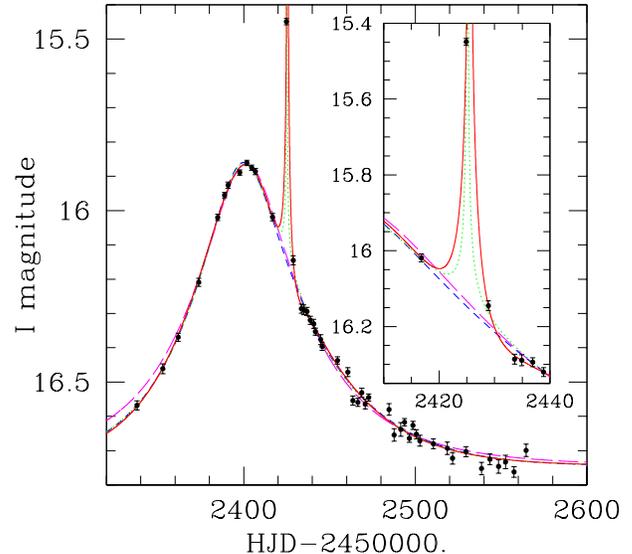}
\caption{\label{fig:one}
The points with errorbars show the light curve of \event.
The long-dashed magenta line shows the best single-lens,
single-source, constant-velocity model fit to the data, with the
single high point and two neighboring points removed.  The
short-dashed blue line shows the best single-lens, single-source,
constant-acceleration model fit to the same dataset.  The dotted green
line shows the best single-lens, binary-source, constant-acceleration
model to the entire data set. The solid red line shows the best
single-source, binary-lens, constant-acceleration model to the entire
dataset.  The inset shows a close-up near the deviation.
}\end{figure}

\section{Data}\label{sec:data}

The Galactic bulge microlensing event \event\ was observed as part of
the third phase of the OGLE collaboration \citep{udalski03b}, and
alerted during the 2002 bulge season with the Early Warning System
(EWS)\footnote{http://www.astrouw.edu.pl/$\sim$ftp/ogle/ogle3/ews/ews.html}.
The OGLE data consist of 106 $I$-band data points, with 9 points taken
during the 2001 season, 47 during the 2002 season, and 50 taken during
the 2003 season.  The data covering the primary event are shown in
Figure \ref{fig:one}.  All errors are scaled by a factor of 1.31, as
determined from the 2003 baseline data.  This scaling may be somewhat
underestimated, as the best models found here have unreasonably large
values of $\chi^2$, although they appear to faithfully reproduce all
the features of the event.  The majority of the data for \event\
appear to follow the usual single-lens form, with the obvious
exception of the data point at ${\rm HJD'}\equiv {\rm
HJD}-2450000.=2424.9$, which has $I=15.449\pm 0.009$, highly
discrepant from the neighboring points at $I\sim 16$ separated by
$4-8$ days.  \citet{jp02} report that this data point is secure, as
evidenced by direct inspection of the raw image.

Unfortunately, there is only $I$-band data available for \event, which
hinders the interpretation of the model fits to the data for several
reasons.  First, without color information during the event, it is
impossible to determine separately the color of the source and blend.
Thus it is not possible to compare the derived source and blend colors
and magnitudes to a local color-magnitude diagram, which provides an
important test of the viability of a given model.  In fact, as this
field was not observed during the second phase of the OGLE
collaboration, there is no $V$-band data available for the source
field at all.  Thus it is also not possible to determine the local $I$-band
extinction $A_I$ via the usual method of `clump-calibration' \citep{ws96,stanek96}.
There are several publicly-available extinction maps for the Galactic
bulge.  The maps of \citet{sumi04} are based on OGLE II data, and thus
do not cover the \event\ field.  \citet{popowski03} have derived
extinction maps based on MACHO collaboration data.  
We estimate the extinction at the location of the source
(${\rm R.A.}=17^{\rm h}59^{\rm m}40.\hskip-2pt^{\rm s}93$, ${\rm
Dec.}=-27^\circ07' 18.\hskip-2pt''2 {\rm (J2000)}$; $l=3.14115,
b=-1.75827$) as the mean of the two closest points (at distances of
$0.12^\circ$ and $0.16^\circ$) in the \citet{popowski03} extinction
maps, weighted by the inverse squared distance.  This yields
$A_V=3.20$, or $A_I=1.57$, where we have adopted $A_I/A_V=0.49$
\citep{udalski03a, sumi04}.

\section{Modeling \event: Technical Details}\label{sec:modelling}

\begin{figure}
\epsscale{2.0} 
\plotone{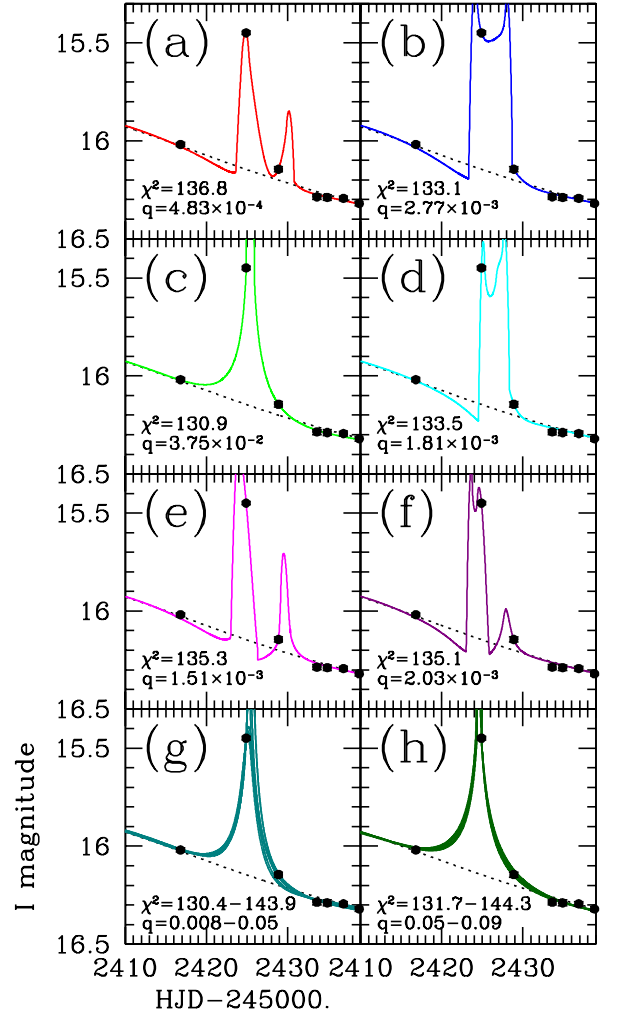}
\caption{\label{fig:two}
In each panel, the points with errorbars show a close-up of
the light curve of \event\ for $\sim 15$ days before and after the
deviant data point.  The solid curve in each panel shows a different
single-source, binary-lens, constant-acceleration fit. The mass ratio
and $\chi^2$ of each fit are also indicated.  The dotted curve shows
the best single-source, single-lens, constant-acceleration fit with
three data points removed, and is the same as the dashed curve in
Figure \ref{fig:one}.  The bottom two panels show a continuum of fits,
with the range of mass ratios and $\chi^2$ values indicated. }
\end{figure}

We fit the flux as a function of time to the standard form,
\begin{equation}
F(t)=\fs A(t)+\fb,
\label{eqn:foft}
\end{equation}
where $\fs$ is the flux of the source, and $\fb$ is the flux of any
unrelated blended light.  Here $A(t)$ is the magnification as a
function of time $t$.

For a single lens (SL) and single source (SS), the magnification is,
\begin{equation} 
A_0[u(t)]=\frac{u^2+2}{u\sqrt{u^2+4}}
\label{eqn:smag}
\end{equation}
where $u(t)$ is the angular separation between the lens and source in
units of the angular Einstein ring radius,
\begin{equation}
\thetae=\sqrt{\kappa M \pirel},\qquad \kappa=\frac{4G}{c^2{\rm AU}}.
\label{eqn:thetae}
\end{equation}
Here $\pirel$ is the lens-source relative parallax, and $M$ is the
mass of the lens.

The angular separation between the source and lens can be written as,
\begin{equation}
u^2(t)=x^2(t) + y^2(t).
\label{eqn:uoft}
\end{equation}
We consider two different models for the relative lens-source motion.
For constant relative velocity (CV) between the source, lens, and
observer,
\begin{equation}
x(t)=\tau\equiv\frac{t-t_0}{\te},\qquad y(t)=u_0
\label{eqn:xycv}
\end{equation}
where $\te=\thetae/\murel$ is the Einstein timescale, $\murel$ is the
relative lens-source proper motion, $t_0$ is the time of closest
approach between lens and source, and $u_0$ is the impact parameter in
units of $\thetae$.  For constant acceleration (CA), the two
components of the angular separation become \citep{smp03},
\begin{equation}
x(t)=\tau + \frac{1}{2}\cax\tau^2,\qquad
y(t)=u_0+\frac{1}{2}\cay\tau^2.
\label{eqn:xyca}
\end{equation}
Here $\cax$ and $\cay$ are the two components of the relative
source-lens angular acceleration in units of $\thetae/\te^2$.  Note
that $\te$ and $\murel$ are now defined at time $t_0$.

For a SL and binary source (BS), the magnification can be written
\begin{equation}
A_{\rm BS}(t)=A_0[u_1(t)]+\epsilon A_0[u_2(t)],
\label{eqn:bsmag}
\end{equation}
where $\epsilon$ is the flux ratio between the secondary and primary
source.  Here $u_1$ and $u_2$ are the angular separations between the
primary and secondary source and the lens, respectively.

For a bound binary-source, we expand the vector angular separation of
the primary and secondary around the time of maximum magnification of
the primary $t_0$, and keep terms up to the acceleration.  This yields
$u_1(t)$ given by \Eq{eqn:xyca}, with $t_0$ and $u_0$ the time of maximum
magnification and impact parameter of the primary,  and $u_2^2(t)=x_2^2(t) + y_2^2(t)$,
with
\begin{equation}
x_2(t)=\tau_1 - \frac{1}{2}\frac{\cax}{q_s}\tau_1^2 +b_x,\qquad
y_2(t)=u_0-\frac{1}{2}\frac{\cay}{q_s}\tau_1^2+b_y,
\label{eqn:xybs}
\end{equation}
where $\tau_1=(t-t_0)/\te$, and $q_s$ is the mass ratio of the binary-source system, and ${\bf b}=(b_x,b_y)$
is the vector angular separation from the primary to the secondary in
units of $\thetae$, which has the same direction as the acceleration
of the primary.  Unfortunately, this parameterization is not ideal for
the current event because $b_x$ and $b_y$ are not directly constrained
by the data, but rather the magnification of the secondary source at
the time of the deviant data point, $A(t_d)\simeq
\epsilon/u_2(t_d)$.  Therefore, adopting \Eq{eqn:xybs} would lead to
correlations between the parameters $\cax, \cay, t_0, u_0$, $b_x$,
and $b_y$, 
which would lead to complications with fitting and error
determination.  Instead, we adopt $x_2(t)=(t-t_{0,2})/\te$ and
$y_2(t)=u_{0,2}$.  This assumes that the acceleration of the secondary
is negligible, and that the direction of the acceleration of the
primary is not necessarily the same as ${\bf b}$.  The acceleration of
the secondary has two effects.  First, the effective timescale at the
time of the perturbation may be different from $\te$.  Second, the
deviation from uniform velocity induces a deviation in the light
curve.  The former effect is essentially unobservable, because the
difference in $\te$ can be absorbed into $\epsilon$ and $u_{0,2}$
\citep{gould96, gaudi98}.  For typical parameters,
the deviation in the light curve
due to acceleration is $\la 1\%$ while the
magnification of the secondary is significant.  This is of order or
smaller than the errors, and thus is negligible.  The alignment
between the source separation vector and acceleration vector as
determined from the fit provides an important test of the
binary-source interpretation.
 
For a binary lens (BL), the magnification is specified by three
parameters: the mass ratio $q$ of the lens, and instantaneous angular
separation $d$ between the two components in units of $\thetae$, and
the angle $\alpha$ between the vector pointing from primary to
secondary and the direction of $\murel$.  Here we adopt a coordinate
system such that the origin of the binary-lens is at a distance of
$qd^{-1}(1+q)^{-1}$ from the primary along the axis toward the
secondary.  For BL models with small mass ratios, it is also necessary
to consider the effect the finite size of the source star $\theta_*$
on the magnification, parameterized by
$\rho_*\equiv \theta_*/\thetae$.

We perform all fits in flux rather than magnitude space, thus allowing
the analytic determination of the values of $\fs$ and $\fb$ for a given
model $A(t)$ that minimize the usual $\chi^2$ goodness-of-fit
statistic.  All other parameters are minimized using the
downhill-simplex routine AMOEBA \citep{numrec92}, using many different
initial trial values of the parameters as seeds.  This procedure works
well for all models except the BL model, which generally produces
poorly-behaved $\chi^2$ surfaces that are not well explored by
downhill-simplex routines.

In order to approximately minimize BL models, and to calculate errors
on fit parameters for other models, we use a Markov Chain Monte Carlo
(MCMC)\footnote{See \citealt{verde03} for a particularly clear and
concise explanation of using MCMC in a practical setting.}.  At each
step in the chain, we vary a single fit parameter by drawing a random
Gaussian deviate with zero mean, and dispersion set to appropriately
sample the likelihood surface.  We then calculate the relative
likelihood ${\cal L}_{rel}=\exp(-\Delta\chi^2/2)$ between the new step,
and the old position. If ${\cal L}_{rel}\ge 1$, the step is accepted.
If ${\cal L}_{rel}< 1$, then we draw a random uniform deviate
between 0 and 1.  If this deviate is $< {\cal L}_{rel}$, then the
new step is accepted, otherwise it is rejected.  When generated in
this way, the resulting distribution of parameters in the Markov Chain
is proportional to the posterior probability distribution, provided that the
chain has converged.  Although we do not rigorously test for convergence, 
we confirm that the chains have been run for a sufficient number of steps,
by visually inspecting different chains started at different points in 
parameter space, and ensuring that these have properly mixed.

We define the $1\sigma$ errors on a given parameter as the projection of the
$\Delta\chi^2=1$ contour onto that parameter axis.  We do not attempt
to calculate errors for the BL models, primarily because we find that
there are large, disconnected regions of parameter space that contain
continua of degenerate models.  Defining confidence regions in the
presence of such degeneracies is difficult and not very meaningful.

\section{Modeling \event: Results}\label{sec:results}

Table 1 shows fit parameters and, where appropriate,
$1\sigma$ errors for the four different classes of models considered
here.  The models are labeled by the lens multiplicity (SL=single
lens, BL=binary lens), source multiplicity (SS=single source,
BS=binary source), and source motion (CV=constant velocity,
CA=constant acceleration).  Also indicated is the dataset used; no
number refers to the entire dataset, `1' indicates the dataset with
the single deviant data point at $t_d=2424.9$ removed, `3' refers the
dataset with points at $t=2416.81$, $2424.9$, and $2428.89$ removed.
For the constant acceleration fits, there are generally always two
degenerate models \citep{smp03}, these are labeled `p' and `n' for the
fits with positive and negative $\cay$, respectively.  The binary-lens
fits are labeled as (a,b,c,d,e,f), which correspond to the panel
labels in Figure \ref{fig:two}.

Following \citet{jp02}, we first fit \event\ to a standard
single-lens, single-source, constant velocity model after removing the
single deviant data point at $t_d=2424.9$.  This model is clearly a
poor fit to the data, yielding $\chi^2\simeq 180$ for 100 degrees of
freedom (dof).  Removing two additional data points on either side of
the deviant data point improves the fit by $\Delta\chi^2=12$ for two
fewer dof.  This implies that the two data points on either side of
the point at $t_d$ likely deviate significantly from the smooth
underlying model, supporting the reality of the single data point, and
pointing toward a perturbation that has a timescale that is of order
the interval between the $t_d$ and neighboring data points, $\sim
4~{\rm days}$.  All constant-velocity fits are quite poor, deviating
significantly from the data during the rising part of the primary
event (see Figure \ref{fig:one}), likely indicating a significant
acceleration of the observer, source, or lens.
 
Including a constant acceleration improves the fits considerably.  For
the data set with one data point removed, we find $\chi^2\simeq 149$
for 98 dof.  When three data points are removed, the fit improves by
$\Delta \chi^2=15.7$ for 2 less degrees of freedom.  As anticipated by
\citet{smp03}, we find two degenerate fits for each constant
acceleration model, with different values of $\te$, $\cax$, and
$\cay$, but identical values of the other parameters and $\chi^2$.
Including the constant acceleration terms increases the inferred
timescale considerably.  The value of $\chi^2$ for the best-fit
constant acceleration model with three data points removed is
unacceptably high, despite the fact that the model appears to
reproduce the primary features of dataset quite well (see Figure
\ref{fig:one}).  Inspection of the light curve during the microlensing
event, while the source is significantly magnified, reveals short
timescale ($\la 1~{\rm day}$) scatter, with amplitude that is smaller
than the typical size of the photometric errors at baseline.  This
scatter may be intrinsic to the source.  Regardless of the cause, it
appears that the error scaling derived from the baseline points is
probably underestimated.

A binary-source model can produce a large, short-timescale deviation
like that seen in the dataset of \event, provided that $\epsilon \ll 1$ and
$u_{0,2}\ll 1$.  Figure \ref{fig:two} shows the best-fit binary-source
models, which have $\chi^2\simeq 142$ for 96 degrees-of-freedom. This
is $\Delta\chi^2\simeq 11$ larger than the single-source
constant-acceleration fit with three data points removed.  This
additional $\chi^2$ arises from the inability of the binary-source
model to simultaneously fit the two data points immediately following
$t_d$.  As with the single-source models, there are two binary-source
models with essentially equal $\chi^2$ due to the
constant-acceleration degeneracy.  As anticipated, the fits yield
small values for the flux ratio, $\epsilon \simeq 1\%$. In this
regime, the relations between the binary-source parameters $\epsilon,
u_0$, and $t_0$ and the salient features of the perturbation are
exceedingly simple \citep{gaudi98}.  Note that the parameters of the
primary ($t_0,\te,u_0,\cax,\cay,\fs, \fb$), which are constrained by
the data away from the perturbation, are essentially identical to the
parameters found for the constant-acceleration fits with three data
points removed. The duration of the perturbation is constrained by the
points neighboring $t_d$.  Combined with the parameters of the
primary, this duration yields $\epsilon$. The excess flux at time
$t_d$ is $\Delta F(t_d)\simeq \fs\epsilon /u_2(t_d)$, and thus yields
$u_2(t_d)$, which is much better constrained than $t_{0,2}$ and
$u_{0,2}$ separately.  We find $u_2(t_d)=0.0025\pm 0.0008$ and
$u_2(t_d)=0.0018\pm 0.0007$ for models SL.BS.CA.n and SL.BS.CA.p,
respectively.  Note that, since the relations between the perturbation
observables and the binary-source parameters $\epsilon, t_{0,2},
u_{0,2}$ depend on $\te$, which differs between the two
constant-acceleration fits, the binary-source parameters also differ
between the two fits.

Binary-lens models with extreme mass ratios $q$ can also yield large,
short-timescale deviations.  In the limit of $q\ll 1$, and under some
simplifying assumptions, the relation between the gross features of
the perturbation and the binary parameters $q,d,\alpha$ also take on a
relatively simple form \citep{gl92,gandg97}.  The angle of the
trajectory is roughly $\alpha\sim \tan^{-1}[{u_0/\tau(t_d)}]$.  The
binary separation is $d \sim 0.5[u_d+\sqrt{u_d^2+4}]$, where $u_d =
\sqrt{\tau(t_d)^2+u_0^2}$.  Finally, the mass ratio is $q\sim
\sqrt{\Delta t/\te}$.  For fits SL.SS.CA.3n and SL.SS.CA.3p, these
expressions yield $\alpha \sim 0.6-0.8$, $d\sim 1.1$, and $q\sim
10^{-3}$.  Of course, these formulas are only very rough
approximations, but the implied parameters are good starting guesses.
We therefore start with these parameter ranges, and first alter the
parameters by hand until we find a number of promising approximate
fits.  These fits are then used as seeds for an automated point-source
binary-lens minimization routine.  All viable point-source fits are
then collected, and the parameters are used as starting guesses for a
second minimization, using a finite source of angular size
\begin{equation}
\theta_*=6~{\mu{\rm as}}~10^{-0.2(I_0-14.32)}.
\label{eqn:thetas}
\end{equation}
Here $I_0$ is the dereddened magnitude of the source as derived from
the point-source fits.  Equation (\ref{eqn:thetas}) is derived from
the color-surface brightness relation of \citet{vb99}, assuming that
the $(V-I)_0$ color of the source is the same as the red clump.
Standard models of the Galaxy \citep{hg95, hg03} predict
$\left<\thetae\right>=240^{+190}_{-130}{\mu{\rm as}}$ (bulge), and
$\left<\thetae\right>=330^{+360}_{-190}{\mu{\rm as}}$ (disk).  These
values of $\thetae$ yield source sizes $\rho_*=\theta_*/\thetae$ that
are too large to reproduce the data for \event\ with small mass ratios
$q$.  Although this implies that such fits are somewhat disfavored, it
is not actually possible to rule out these models with such an
argument.  We therefore adopt a value of $\thetae=550{\mu{\rm as}}$.
Approximately $6\%$ of bulge lenses and $25\%$ of disk lenses are
expected to have values of $\thetae$ larger than this.

Representative finite-source, binary-lens fits are tabulated in Table
1 and shown in Figure \ref{fig:two}.  We recover the fits
presented by \citet{jp02}, and find many other additional viable fits,
with mass ratios ranging from $q\simeq 5\times 10^{-4}$ to $\sim
8\times 10^{-2}$, at the $\Delta \chi^2\la 9$ level.  The best
binary-lens (and best overall) fit has $\chi^2=130$ for 96 dof.  If
one forces $\chi^2/{\rm dof}=1$ for this fit, then mass ratios in the
range $q=10^{-4}-10^{-2}$ are all consistent with the data at the
$\sim 3\sigma$ level.

There two basic classes of fits.  In the first class the trajectory
crosses the planetary caustic, with the deviant data point at $t_d$
occurring when the source is interior to, or near, the caustic.  The
second class of fits pass outside the planetary caustic, with the
point at $t_d$ occurring when the source is near the ridge of
high-magnification on the planet-star axis.  The second class actually
represents a continuous degeneracy in $q,d$, and $\alpha$, with mass
ratios in the range $0.01\la q \la 0.1$ for $\Delta\chi^2\la 14$.
From Figure \ref{fig:two}, we conclude that while many of the fits
would have been distinguishable with a few additional data points
during the perturbation, the family of continuous fits produce very
similar perturbations.  The curves in each of the two bottom panels of
Figure \ref{fig:two} deviate from each other by $\la 2\%$ for the
majority of the perturbation.  Distinguishing between these models
would therefore require rather accurate or densely sampled data.  
The cause of this uncertainty in $q$ is briefly discussed 
by \citep{gandg97}.

\section{Implications and Discussion}\label{sec:implications}

\event\ can be reasonably well-fit by large number of different
models.  \citet{jp02} demonstrated that the asymmetry exhibited by the
primary light curve can be explained by a simple uniform acceleration,
due to either parallax deviations arising from the motion of the
Earth, or non-uniform motion of the source due to the presence of,
e.g., a binary companion.  We have demonstrated that the
short-timescale deviation can be explained by a binary companion to
the lens with mass ratio $10^{-4}\la q \la 10^{-1}$, consistent with a
planet, brown dwarf, or low mass star.  The short-timescale deviation
can also be explained by a binary companion to the source with a flux
ratio of $\sim 1\%$.

Which of these models is most likely?  \citet{jp02} showed that the
parallax and constant acceleration fits had essentially identical
$\chi^2$-values for the same number of degrees-of-freedom.  Thus
either are equally viable in a goodness-of-fit sense.  The inferred
timescale of the event spans the range $\te=50-100~{\rm days}$, which
is in the range where weak parallax effects are neither surprising,
nor necessarily expected.  Furthermore, there has been no estimation
of the expected rate of binary-source events with detectable
asymmetries due to acceleration.  Therefore it is difficult to argue
which origin for the observed asymmetry is a priori more likely.  In
regards to the deviant data point, a binary-lens model with $q\sim
{\rm few}\times 10^{-2}$ provides the best fit to the data, with the
binary-source model disfavored at $\Delta \chi^2=11$ for the same
number of dof, or slightly more than $\sim 3\sigma$ level.  However,
this difference in $\chi^2$ is driven primary by a couple of data
points just after the perturbation, which may be affected by the
short-timescale variability seen in other parts of the light curve.
If one normalizes the error bars to force $\chi^2/{\rm dof}=1$ for the
best model, then the binary-lens model is only favored at the
$\Delta\chi^2\simeq 8$ level.  The inferred blend and source fluxes can
also provide discrimination between models: one expects these fluxes
to trace, on average, the distribution of fluxes of unlensed stars in
the field near the line-of-sight.  The best binary-lens models have
source and blend magnitudes of $I_s\sim 17$ and $I_b\sim 19$, whereas
the binary-source fits yield $I_s \sim 18$ and $I_b\sim 17$.  
Unfortunately, without color information, it is 
difficult to definitively distinguish between these two scenarios.
It is also difficult
to argue which model is a priori more likely, since the frequency of
planetary companions to microlens stars is not known, and defining the
appropriate detection criterion for the current event, which was
culled by eye from an ensemble of alerted microlensing events, is
nebulous at best.  

The binary-source model does have the advantage of simplicity: a bound
companion to the primary source can explain both the short-timescale
variability and the acceleration needed to produce the global
asymmetry.  However, in order for this scenario to hold, the projected
acceleration vector must be aligned with the projected binary-source
separation vector.  In the particular parameterization of the
binary-source model used here (see \S\ref{sec:modelling}), this angle
$\Delta \theta$ between the two vectors is essentially a free
parameter, and thus it can be used to test the model.  We determine
the distribution of $\Delta \theta$ from the two Markov chains
corresponding to the two degenerate (in the sense of $\chi^2$)
binary-source models, SL.BS.CA.n and SL.BS.CA.p.  The second model is
immediately ruled out, as the primary's acceleration vector is
pointing away from the companion, with $\Delta \theta>90^\circ$,
wildly inconsistent with any bound orbit.  The first model has $\Delta
\theta = 36.2_{-2.0}^{+1.8}$ degrees.  At first sight, this may also seem very
inconsistent ($>15\sigma$!) with a bound orbit ($\Delta \theta=0$),
however there are several reasons why this is not necessarily correct.
First, although the $1\sigma$ uncertainty on $\Delta\theta$ is small,
there is a substantial non-Gaussian tail toward smaller values.
Enforcing $\Delta\theta=0$ in the fit yields $\chi^2=156.7$, or
$\Delta\chi^2=15.0$ for one less dof.  Thus, the bound model is only
`ruled out' at the $\sim 3.5\sigma$ level.  Furthermore, as we argue
below, the period of a bound binary source with separation $b\la 1$ is
likely to be only a few times longer than the timescale of the event.
Therefore one expects significant deviations from uniform acceleration
that are not accounted for in the simplified model adopted here.  Thus
the inferred value of the $\Delta\theta$ may be the result of
inadequacies in the model.  It is therefore conceivable \event\ could
be fit by a bound binary-source model in which the effects of rotation
are included self-consistently.

Accepting the binary-source model as correct, one can use the measured
parameters of the viable model SL.BS.CA.n to determine the Einstein
ring radius projected on the source plane $\rehat=\dos\thetae$
\citep{hg97}, where $\dos$ is the distance to the lens.  Assuming
circular orbits, the acceleration of the primary due to the secondary
is simply $a=G m_2/d^2$, where $d$ is the semi-major axis of the orbit
and $m_2$ is the mass of the secondary.  The projected acceleration
$a_\perp$ is therefore,
\begin{equation}
a_\perp=\frac{G m_2}{d^2}(1-\sin^2{\phi}\sin^2{i})^{1/2}
\label{eqn:aperp}
\end{equation}
where $i$ is the inclination of the orbit, and $\phi$ is the phase of
the primary orbit referenced to the intersection of the orbit and the
plane of the sky.  The constant-acceleration fit yields the parameter
${\cal A}=\sqrt{\cax^2+\cay^2}$, which is given by,
\begin{equation}
{\cal A}=\frac{a_\perp \te^2}{\rehat}.
\label{eqn:accdim}
\end{equation}
The binary-source fit yields $t_{0,2}$.  This can be combined with the
parameters of the primary event to derive the two components the
projected binary-source separation:
$b_x=0.5(\cax/q_s)\Delta\tau^2-\Delta\tau$ and
$b_y=0.5(\cay/q_s)\Delta\tau^2-u_0$.  Here $\Delta
\tau\equiv(t_{0,2}-t_0)/\te$, and we have assumed that $u_{0,2}\ll 1$.
Then $b$ can be related to the semi-major axis by,
\begin{equation}
d=\rehat b(\cos^2{\phi}+\cos^2{i}\sin^2{\phi})^{-1/2}.
\label{eqn:dtob}
\end{equation}
Finally, Equations (\ref{eqn:aperp}), (\ref{eqn:accdim}), and
(\ref{eqn:dtob}) can be combined to arrive at an expression for
$\rehat$ in terms of mostly known quantities,
\begin{equation}
\rehat=\left[\frac{G m_2 \te^2}{b^2 {\cal A}}
(1-\sin^2{\phi}\sin^2{i})^{1/2}(\cos^2{\phi}+\cos^2{i}\sin^2{\phi})\right]^{1/3}.
\label{eqn:rehat}
\end{equation}

We now use the probability distribution of the binary-source model fit parameters,
together with \eq{eqn:rehat}, to construct the probability
distribution for $\rehat$.  Here the virtue of MCMC becomes
clear: in order to determine the probability distribution
of $\rehat$, we simply have to determine the value of $\rehat$ at each link in
the chain.  Because the density of points in the binary-source MCMC parameter chain
is proportional to the posteriori probability distribution, the resulting
distribution of values of $\rehat$ is simply proportional to the desired probability
distribution.  

We first account for the parameters 
in \eq{eqn:rehat} that are not constrained by the binary-source model;
these include the inclination $i$, the phase $\phi$, and the mass ratio $q_s$ (which 
enters via $b$).  For each link in the chain, we
randomly choose ten values of $\phi$ and $\cos{i}$, assuming a uniform
distribution of each.  We determine $q_s$ using the 
fluxes of the primary and the secondary.  We draw a source distance $\dos$ from the
standard Galactic model of \citet{hg95,hg03}.  If the source
is in the bulge, we assume that the primary is a giant, and therefore
has a mass of $M_1\sim M_\odot$.  If the source is in the disk, we determine
the absolute magnitude of the primary from the value of $\dos$, and
assuming an extinction of $A_I=1.57$ (see \S\ref{sec:data}).  We determine
the absolute magnitude of the secondary from 
the flux ratio $\epsilon$, primary flux $F_s$, $\dos$, and $A_I$.   We
estimate the masses of the
primary and secondary from their absolute magnitudes
using an analytic approximation to the main-sequence mass-luminosity
relationship derived from the 
solar metallicity isochrones of \citet{bertelli94}.  This then yields the mass ratio $q_s$.  We then determine $\rehat$.

In addition, we also determine the radius $R_{*,2}$ of the secondary
from its absolute magnitude, also using the solar-metallicity
isochrones of \citet{bertelli94}.  
We require
that $\rho_{*,2}\equiv R_{*,2}/\rehat<u_2(t_d)$, otherwise we discard
the inferred value of $\rehat$.  We note that the resulting
distribution of $\rehat$ is not very sensitive to the adopted
distribution of source distances, for several reasons.  First, the
distance enters through the factor of $b^{-2/3}$, since $b$
formally depends on the mass ratio $q_s$.  However, inspection of the
expressions for $b_x$ and $b_y$ reveals that $q_s$ modifies the
acceleration terms, which themselves are $\propto \tau_2^2$,
where $\tau_2=(t_{0,2}-t_0)/\te$.   For the
binary-source fits presented here, $\tau_2^2\sim 10\%$, and thus these
terms are small.  Second, for bulge sources, where $q_s$ is affected the
most by the assumed source distance, the dispersion in source
distances is relatively small.  Conversely, for disk sources, where
the dispersion in source distances is large, the inferred value of
$q_s$ is relatively insensitive to the source distance (since both
primary and secondary are assumed to be on the main sequence). 
Had we simply assumed that all sources were in the bulge
with $\dos=8~{\rm kpc}$, the median inferred value of $\rehat$ 
would change by $\la 2\%$, or $\sim 0.1\sigma$. 

\begin{figure}
\epsscale{2.0} 
\plotone{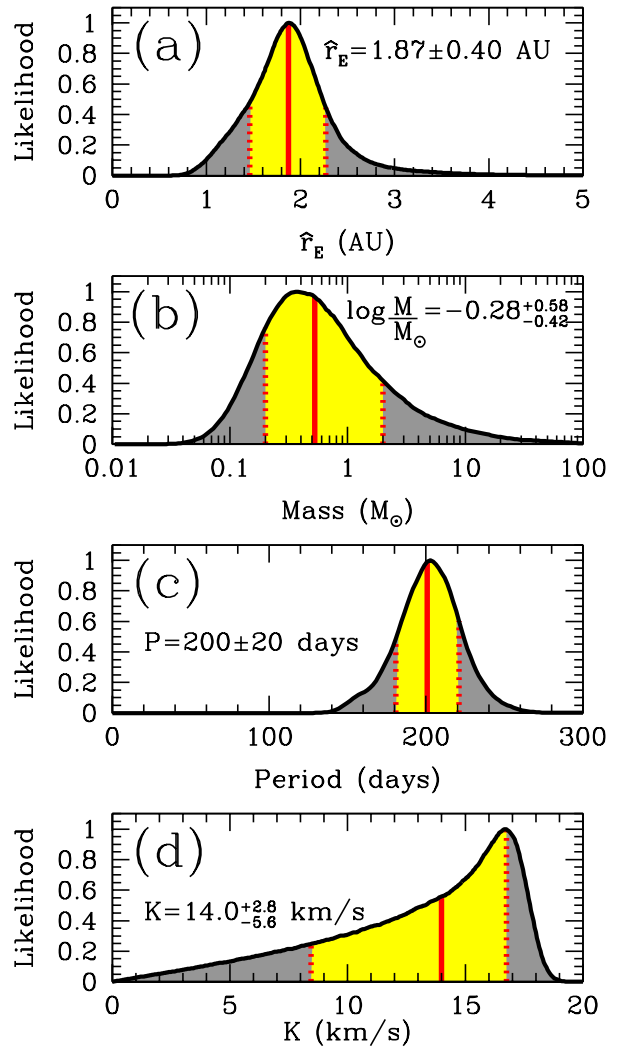}
\caption{\label{fig:three}In each panel, the curves show the relative likelihood of
the given parameter, as inferred from binary-source, constant
acceleration fit SL.BS.CA.n (see Table 1 and
\S\ref{sec:implications}).  All curves have been normalized so that
the likelihood is unity at the peak.  The solid vertical line
shows the median of the distribution, while the dotted lines show the
68\% confidence limits. (a) The relative likelihood of $\hat r_{\rm
E}$, the Einstein ring radius of the lens projected on the source
place.  (b) The relative likelihood of the mass of the lens, as
inferred from the measured values of $\rehat$, the
timescale $\te$, and the standard Galactic model of
\citet{hg95,hg03}.  (c) The relative likelihood of the period of the
binary-source in days.  (c) The relative likelihood of the
semi-amplitude $K$ of the radial velocity of the primary of the binary
source, in kilometers per second.}
\end{figure}

Figure \ref{fig:three} shows the
resulting distribution of $\rehat$.  The median and 68\% confidence
interval is $\rehat=1.87\pm0.40~{\rm AU}$.  
Adopting the standard Galactic model of \citet{hg95,hg03}, we use the inferred
values of $\rehat$ and $\te$ to constrain the mass $M$, distance $\dol$
and relative proper motion $\murel$ of the lens.
The resulting distribution of $M$ (assuming a uniform prior in linear
mass) is shown in Figure \ref{fig:three}.  The median and $68\%$
confidence interval is $\log{M/M_\odot}=-0.28_{-0.42}^{+0.58}$. 
Similarly, we infer $\dol=7.1\pm 0.9~{\rm kpc}$, $\thetae=240_{-40}^{+45}\mu{\rm as}$, and 
$\murel=5.45_{-0.95}^{+1.15}~{\rm km~s^{-1}~kpc^{-1}}$.  All
of these parameters are typical for disk-bulge or bulge-bulge lensing
events, except for $\murel$, which is a factor of $\sim 6$ smaller
than the median distribution for bulge-bulge events and a factor of
$\sim 8$ smaller than the median of disk-bulge events.  This may be
due to the fact that the lens and source happen to have usually small
relative velocities, it may indicate that the event is due to
disk-disk lensing, or it may simply be because the model is incorrect.

We also infer an Einstein ring radius projected onto observer plane of
$\tilde r_{\rm E}=13.3_{-7.0}^{+15.5}~{\rm AU}$, indicating that, for
the binary-source model, parallax effects are likely to be negligible.

The probability distributions of several other interesting quantities
can also be constructed from the Markov chain.  The semi-major axis of
the binary is $d=0.80\pm0.06~{\rm AU}$.  In order to determine the
period, we assume a primary mass of $m_1=M_\odot$, as expected for
bulge giants.  The absolute $I$-band magnitude of the secondary is
$M_I=6.60\pm 0.46$, which corresponds to a mass of $m_2=0.67\pm 0.05~M_\odot$
assuming that it is on the main-sequence.  These corresponds to a late
K-dwarf.  The primary has an absolute magnitude of $M_I=2.01\pm 0.49$,
making it likely to be a $M_1\sim M_\odot$ bulge giant.  The total
binary mass is $M_{s}\equiv m_1+m_2 \simeq 1.7M_\odot$.  The
distribution of the binary period is shown in Figure \ref{fig:three},
the median and 68\% confidence interval is $P=200\pm 20~{\rm days}$.
The binary period is only a factor of $\sim 2$ larger than the
microlensing event timescale $\te$.  This implies that the assumption
of a uniform acceleration is almost certainly violated, and that the
binary-source model is not internally self-consistent.  The constant
acceleration approximation may not be so bad, as the deviations from
uniform acceleration will only become significant in the tails of
the event, when it is not highly magnified.  Nevertheless, the
parameters derived from the above analysis should be interpreted with
caution.

It is also possible to predict the semi-amplitude of the radial
velocity of the primary,
\begin{equation}
K=\left[\frac{Gm_2^2}{dM_{\rm s}}\right]^{1/2}\sin{i}.
\label{eqn:k}
\end{equation}
The distribution of $K$ is shown in Figure \ref{fig:three}.  The
median and 68\% confidence interval is $14.0_{-5.6}^{+2.8}~{\rm
km~s^{-1}}$.  Radial velocity precisions of a few ${\rm km~s^{-1}}$
should be attainable for this source, which has $I=17.9$.  Thus it may
be possible to directly confirm the binary-source interpretation of
this event.

\section{Conclusion}

Microlensing event \event\ exhibits a slightly asymmetric, smooth
light curve with a single data point that deviates by $\sim 0.6$
magnitudes from neighboring points separated by several days.
\citet{jp02} concluded that the simplest interpretation of \event\ was
an event with parallax deviations arising from the motion of the
Earth, and a short-timescale deviation due to a binary lens with mass
ratio $q=10^{-3}-10^{-2}$, thereby making \event\ a candidate
planetary event.

Here we demonstrated that \event\ can be reasonably well-fit by
several different classes of models, and a wide range of parameters
within each model class.  We found that the data can be fit by many
different binary-lens models whose mass ratios span three orders of
magnitude, from $q=10^{-4}$ to $10^{-1}$, thereby making the secondary
consistent with a planet, brown dwarf, or M-dwarf for reasonable
primary masses.  A subset of these binary-lens fits form a family of
continuously degenerate models, whose mass ratios differ by an order
of magnitude.  Astonishingly, the light curves of these models differ
by $\la 2\%$ for the majority of their duration.  A binary-source
model is also consistent with the data, for a secondary/primary flux
ratio of $\sim 1\%$.  This model also naturally explains the global
asymmetry of the lightcurve as due to the acceleration of the primary
induced by the secondary.  Under the assumption of a bound
binary-source, this model yields an estimate of the Einstein ring
radius projected on the source plane of ${\hat r_{\rm E}}=1.87\pm 0.40~{
\rm AU}$.

All of these fits differ by $\la 3\sigma$, and are essentially
indistinguishable when the scatter due to likely source variability is
considered.  Unfortunately, the lack of color information during the
event precludes the discrimination of models based on the positions of
the source and blend on a color-magnitude diagram.

Although the primary goal of the study by \citet{jp02} was to affect a
modification of the OGLE observation strategy to ensure good sampling
of short-duration perturbations, rather than argue that \event\ was a
bona fide planetary event, it is still somewhat disturbing that many
binary-lens fits were missed, and the possibility of a binary-source
interpretation was not discussed at all.  \event\ serves as an extreme
reminder of the degeneracies inherent in microlensing events, and
highlights the difficulties in interpreting poorly-sampled and
weakly-perturbed events. These difficulties become especially
important when attempting to detect planets with microlensing.  Here
observers and modelers need to be especially vigilant: in order to
produce convincing and reliable planetary detections, it is essential
not only to achieve dense and accurate photometry of planetary
perturbations, but also to acquire as much auxiliary information as
possible, and to perform detailed, careful, and thorough modeling, in
order to ensure that planetary detections are robust.

\acknowledgments We would like
to thank the anonymous referee for useful comments and suggestions, 
and the OGLE collaboration for making its microlensing data
publically available.  
This work was supported by a Menzel Fellowship from
the Harvard College Observatory, by the National Science Foundation
under Grant No.\ PHY99-07949, by the Astrophysical Research Center for
the Structure and Evolution of the Cosmos (ARCSEC) of the Korean
Science \& Engineering Foundation (KOSEF), through the Science
Research (SRC) program, and by JPL contract 1226901.
B.S.G.\ would like to acknowledge the hospitality of the staff and
participants of the workshop on ``Planet Formation: Terrestrial and
Extra Solar,'' at the Kavli Institute for Theoretical Physics,
University of California, Santa Barbara, where the majority of this
work was completed.

\begin{figure}
\epsscale{1.2} 
\plotone{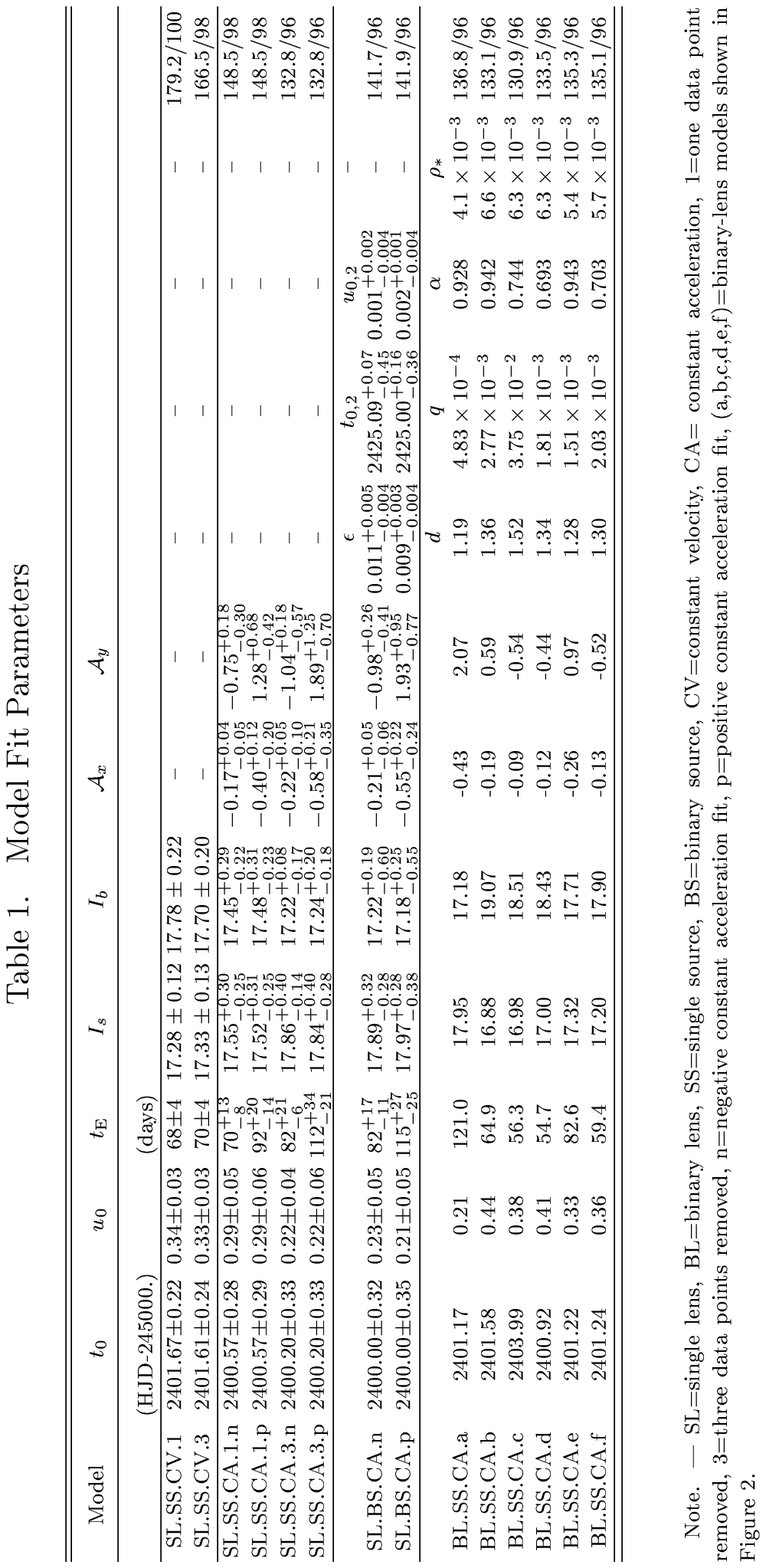}
\end{figure}

\end{document}